      [1999/12/01 v1.4c Il Nuovo Cimento]
\documentclass{cimento}
\usepackage{graphicx}  

\def\c2nor{\chi^2}

\def\epi{$E_{\rm p,i}$}
\def\eiso{$E_{\rm iso}$}

\def\epeiso{$E_{\rm p,i}$ -- $E_{\rm iso}$}

\def\nufnu{$\nu$F$_\nu$ }

\def\swift{{\it Swift}}

\title{\swift{} GRBs and the \epeiso{} correlation} 
\author{L.~Amati}
\instlist{\inst INAF - IASF Bologna, via P. Gobetti 101, Bologna (Italy)} 
\PACSes{\PACSit{98.70.Rz}{gamma--ray sources; gamma--ray bursts}
}
\begin{document}

\maketitle

\begin{abstract}
The \epeiso{} correlation is one
of the most intriguing and debated observational evidences in Gamma-
Ray Bursts (GRB) astrophysics. \swift, with its high sensitivity and fast
pointing capabilities, is reducing a lot the impact of selection effects in the
sample of GRBs with known redshift (and thus \epi{} and \eiso). Moreover, in
several cases it allows the detection of the soft tail of the prompt emission, and
thus a more accurate estimate of \epi{} with respect to previous satellites.
I present
and discuss the location in the \epeiso{} plane of \swift{} GRBs with known 
redshift and estimated \epi, showing that all long events (inlcuding peculiar
events like GRB\,060218 and GRB\,060614) are consistent with
the \epeiso{} correlation. In contrast, short GRBs are
not consistent with it, an evidence further supporting the hypothesis of
different emission mechanisms at work in the two classes of GRBs.
I also show, and briefly discuss, the intriguing evidence that
the soft tail of the short GRB\,050724 is consistent with
the correlation.
\end{abstract}

\section{Introduction}

In the last decade, thanks to the discovery and study of afterglow emission and
host galaxies, it has been possible to estimate the redshift of several
tens of Gamma--Ray Bursts (GRBs), and thus to derive
their distance scale, luminosities and 
other intrinsic properties.
Among these, the correlation between the cosmological rest--frame \nufnu spectrum peak
energy, \epi, and the isotropic equivalent radiated energy, \eiso,
is one of the most intriguing and robust. Indeed, as shown, initially by 
\cite{Amati02,Amati03,Ghirlanda04a,Friedman05} and, more recently, 
by \cite{Amati06}, all
long GRBs with known redshift and estimated \epi{} are consistent with
the relation \epi{} = K $\times$ $E_{\rm iso}^{m}$ 
(K $\sim$75--110 and $m$ $\sim$0.4--0.6, with \epi{} in keV and
\eiso{} in units of 10$^{52}$ erg),
with the only exception of GRB\,980425 (which is anyway
a peculiar event under several other aspects). The \epeiso{} correlation holds
from the brightest GRBs to the weakest and softest X--Ray Flashes (XRFs)
and is characterized by a scatter in log(\epi) of $\sim$0.2 dex
(by assuming a 
Gaussian distribution of the deviations).
The implications and uses of the \epeiso{} correlation include
prompt emission physics, jet geometry and structure, testing of GRB/XRF synthesis and
unification
models, pseudo--redshift
estimators, cosmology (when additional observables, like e.g. the break time of the 
optical afterglow light curve or the high signal time scale, are included
\cite{Ghirlanda04a,Liang05,Firmani06}); see \cite{Amati06} for a review. 

In the recent years
there has been a debate, mainly based on BATSE GRBs without known redshift,
about the impact of selection effects on the 
sample of GRBs with known redshift and thus on the \epeiso{} correlation. 
Based on the analysis of BATSE GRBs without known redshift, different conclusions 
were reported (e.g., \cite{Nakar05,Band05,
Ghirlanda05,Pizzichini06}), but with the general agreement that 
\swift{} would allow us to test the correlation in a more stringent way:
\begin{itemize}
\item{with respect, e.g., to BATSE, the BAT GRB detector has a sensitivity
which is comparable
for GRBs with peak energies above $\sim$100 keV, but much better for softer
events (e.g., \cite{Band03}}), thus reducing selection effects at the detection stage;
\item{the fast pointing capability permits few arcsec localizations with XRT and their
dissemination to optical telescopes as early
as $\sim$100-200s form GRB onset, thus reducing selection effects at the redshift 
estimate stage; 
this is clearly demonstrated by the fact the redshift estimates for \swift{} GRBs are
more frequent 
and differently distributed with respect to the past 
(see, e.g., \cite{Berger05,Jakobsson06}).
} 
\end{itemize}

In addition, thanks to the fast pointing capability, it is possible in some cases, 
to follow the later
and softer phase of the prompt emission with XRT, thus providing a more
accurate estimate of the peak energy (as in the case of GRB\,060218
\cite{Campana06}).
From the point of view of testing the \epeiso{} correlation, the only negative
of \swift{} is that BAT, due to its limited energy band ($\sim$15--350 keV),
provides an estimate the spectral peak energy only for a small fraction (15--20\%) 
of the events.
Fortunately, this is partially compensated by the simultaneous detection of several
\swift{} GRBs by broad band experiments like HETE--2, Konus/Wind and, more recently,
RHESSI and Suzaku/WAM, all capable to provide estimates of \epi{} for most
of the events detected by them.

\begin{table}
\caption{\epi{} and \eiso{} values for \swift{} GRBs/XRFs with known redshift and firm 
estimates of (or upper/lower limits to) \epi. All values have been taken from the literature or 
computed based on published spectral parameters and fluences, see last
column for references (an asterisk indicate that the references can be found
at http://swift.gsfc.nasa.gov/docs/swift/archive/grb\_table/).
For GRB\,050724, the first line corresponds to the short pulse, the second
line to the soft tail.
}
\begin{tabular}{llllllc}
\hline
 GRB  & Type & $z$  & \epi{}  & \eiso{} & Instruments &  Refs.  \\ 
      &   &   &  (keV)     & (10$^{52}$ erg)  &  &  \\
\hline
      050315  &  LONG &     1.949 &  $<$89 &  4.9$\pm$1.5
 &       SWI &   \cite{Amati06}  \\
      050318  &  LONG &     1.44 &  115$\pm$25 &  2.55$\pm$0.18
 &       SWI &   \cite{Amati06} \\
      050401  &  LONG &     2.90 &  467$\pm$110 &  41$\pm$8 & 
      KON &  * \\
      050416a & LONG &     0.650 &  25.1$\pm$4.2 &  0.12$\pm$0.02
 &       SWI &  \cite{Amati06}  \\
      050505 & LONG &    4.27 &  $>$274 &  40$\pm$10
 &       SWI &  * \\
      050509b  &   SHORT &   0.22 &  $>$183 &  0.0007$\pm$0.0004
 &       SWI & * \\
      050525  &   LONG &   0.606 &  127$\pm$10 &  3.39$\pm$0.17
 &       SWI & \cite{Amati06}  \\
      050603  &   LONG &    2.821 &  1333$\pm$107 &  70$\pm$5 & 
      KON & \cite{Amati06}  \\
      050724 &   SHORT &   0.258 &  $>$126 &  0.03$\pm$0.01
 &       SWI & * \\
      050724  &   SHORT &   0.258 &  11$\pm$2 &  0.035$\pm$0.007
 &       SWI & * \\
      050730  &   LONG &   3.967 &  $>$750 &  26$\pm$19
 &       SWI & \cite{Perri06}  \\
      050813  &   SHORT &   0.72 &  $>$344 &  0.09$\pm$0.06
 &       SWI & * \\
      050824  &   LONG &   0.83 &  $<$23 &  0.13$\pm$0.029
 &       HET & \cite{Amati06} \\
      050904  &   LONG &   6.29 &  $>$1100 &  193$\pm$127
 &       SWI & \cite{Amati06}   \\
      050922c  &  LONG &     2.198 &  415$\pm$111 &  6.1$\pm$2.0 & 
      HET &  \cite{Amati06}  \\
      051022  & LONG &  0.80 &  754$\pm$258 &  63$\pm$6 & 
      HET/KON &  \cite{Amati06} \\
      051109  & LONG &      2.346 &  539$\pm$200 &  7.5$\pm$0.8
 &       KON &  \cite{Amati06}   \\
      051221a  &  SHORT &   0.5465 & 622$\pm$35  &  0.29$\pm$0.06 
 &       KON &  \cite{Amati06}   \\
      060115  & LONG &    3.53 &  281$\pm$93 &  9.1$\pm$1.5
 &       SWI &  * \\
      060124  & LONG &   2.296 &  784$\pm$285 &  48$\pm$7
 &       KON &  * \\
      060206  & LONG &      4.048 &  380$\pm$95 &  5.8$\pm$0.6
 &       SWI &  * \\
      060218  & LONG &      0.0331 &  4.9$\pm$0.3 &  0.0062$\pm$0.0003
 &       SWI &  \cite{Amati06b}   \\
      060418  & LONG &      1.489 &  572$\pm$143 &  15$\pm$3
 &       KON &  * \\
      060502b  & SHORT &    0.287 &  $>$193 &  0.025$\pm$0.020
 &       SWI &  * \\
      060505 & SHORT?&  0.089 &  $>$160 &  0.003$\pm$0.001
 &       SWI &  \cite{Amati06b}   \\
      060614  & LONG &   0.125 &  55$\pm$45 &  0.25$\pm$0.1
 &       KON &  \cite{Amati06b}   \\
      060707  & LONG &      3.425 &  290$\pm$27 &  8.0$\pm$1.5
 &       SWI &  * \\
      060927  & LONG &   5.6 &  470$\pm$120 &  10$\pm$2
 &       SWI &  * \\
      061007  & LONG &    1.261 &  890$\pm$124 &  100$\pm$10
 &       KON &  \cite{Mundell06}   \\
\hline
\end{tabular}
\vspace{-0.8cm}
\end{table}

\section{Long GRBs}

The sample of \swift{} long GRBs with known redshift and published spectral peak energy
consists of 17 events; for other 5 events upper/lower limits to \epi{} have been reported.
These events are listed in Table 1; 
for $\sim$half of the events, the values of (or upper/lower limits to) \epi{} and \eiso{} are taken from \cite{Amati06}; for
the others, they have been computed based on published spectral information 
following the method reported in \cite{Amati02,Amati06} and
by assuming a cosmology with $\Omega_m$ = 0.3, $\Omega_{\Lambda}$ = 0.7 and H$_0$ =
65--70 km s$^{-1}$ Mpc$^{-1}$ (i.e. the uncertainties on \eiso{} take into account the
uncertainty in the value of H$_0$). As can be seen in Figure 1, all these events (filled circles) are fully consistent with the \epeiso{} correlation as characterized by the best fit 
power--law and logarithmic scatter derived by \cite{Amati06}. The fit with a 
power--law of the \swift{} sample of 17 long GRBs with firm estimates of $z$ and \epi{}
gives an index of 0.58$\pm$0.01 and a normalization of 86$\pm$3 (\epi{} in keV and
\eiso{} in units of 10$^{52}$ erg ). Consistently again with the findings based on
data from previous 
satellites, despite the correlation is very significant (the Spearman's rank correlation 
coefficient between log(\epi{}) and log(\eiso) of the 17 \swift{} events is $\sim$0.93), the
chi--square value obatined with the power--law fit is high (47/15), confirming the
presence of extra--Poissonian dispersion. As can be seen in Figure 1, 
the scatter of the data in terms of log(\epi{}) around the best-fit power--law can be
fitted by a Gaussian with $\sigma$$\sim$0.16, slightly lower than the $\sim$0.2 found by
\cite{Amati06}.
Finally, as can be seen in Figure 1, peculiar \swift{} events like the sub--energetic, 
very close GRB\,060218/SN2006aj and 
GRB\,060614, a long GRB not associated with a hypernova, are both consistent 
with the correlation,
an evidence that gives important clues for the understanding of their
nature (as shown and discussed by
\cite{Amati06b}).

\begin{figure}
\centerline{\includegraphics[width=11cm]{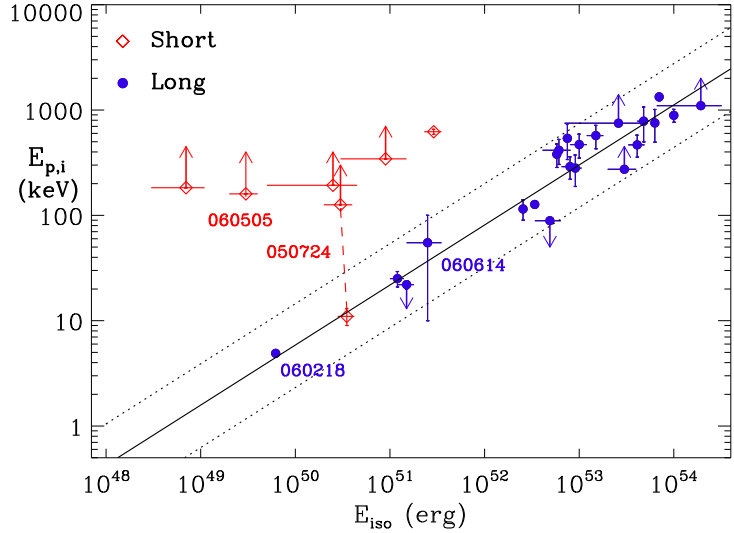}}
\caption{Location in the \epeiso{} plane of \swift{} long (filled circles) and short (diamonds)
GRBs with known redshift and available estimates of \epi. The long GRBs sample 
includes also 4 GRBs for which upper/lower limits to \epi{} have been reported; the lower limits to \epi{} for short GRBs have been estimated based on the BAT spectral
photon indices (see text).
The continuous line is the
power--law best fitting the \epeiso{} correlation and the dashed lines
delimitate the 2$\sigma$ confidence region (from \cite{Amati06}).}
\end{figure}

\begin{figure}
\centerline{\includegraphics[width=7.3cm]{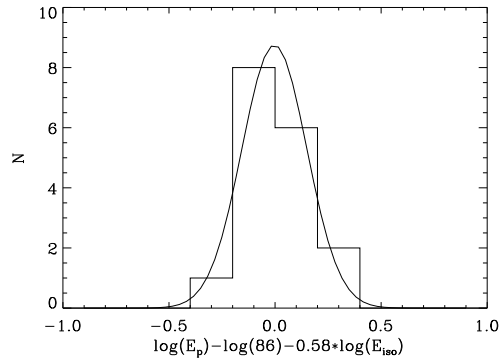}}
\caption{Dispersion of the values of log(\epi) of 17 long Swift GRBs with respect to the
power--law best fitting the \epeiso{} correlation, modeled with a Gaussian 
($\sigma$$\sim$0.16).
}
\end{figure}

\section{Short GRBs}

An important breakthrough of the \swift{} era is the discovery of afterglow emission from
short GRBs, leading to the first redshift estimates (starting with GRB\,050709 detected
by HETE--2) for this elusive events. Even if short GRBs with known redshift are still
a few, there is evidence that, as long GRBs, they lie at comsological distances, even if 
at lower redshifts ($<$$\sim$0.7). As can be seen in Table 1 and Figure 1, only for 1 
of \swift{} 
short GRBs (051221) there is an estimate of \epi; the other short GRB with known
redshift and \epi{} is GRB\,050709. As already shown by \cite{Amati06}, both these
events are inconsistent with the \epeiso{} correlation (see Figure 3). For the other
\swift{} short GRBs, also reported in Table 1 and shown in Figure 1 (diamonds), only
{\it approximate} lower limits to \epi{} can be inferred based on the photon indices estimated by 
fitting
the BAT spectra with a power--law. In all cases, the photon index is hard enough to
support the hypothesis of a peak energy at least higher than 100 keV (the fits of BAT spectra
reported in GCNs are typically performed in the 15-150 or 15--350 keV energy band).
More specifically, based on the reported photon indices and energy bands, the lower 
limit to the observer's frame peak energy was chosen to be 100 keV for GRB\,050724,
150 keV for GRBs 050509b, 0605002b and 060505, 200 keV for GRB\,050813.
For each event,  \eiso{} was computed by assuming \epi{} varying between its lower limit and
(conservatively) 10000 keV. Despite its T$_{90}$ of 4$\pm$1 s, I included GRB\,060505
in the short GRB sample, because of its very low fluence and hard spectrum (typical 
features of short GRBs) and its duration anyway consistent with the tail of short
GRBs duration distribution (see also \cite{Amati06b}).
In Figure 1 it can be seen that all \swift{} GRBs with known redshift
are inconsistent with the \epeiso{} correlation; in particular, they lie significantly
above the region populated by long events. 
As discussed by \cite{Amati06}, the different location of long and short GRBs in the
\epeiso{} plane is consistent with the different distributions of
these two classes in the hardness--intensity diagram and can give important
clues for the understanding of the differences in their emission mechanism(s) 
and progenitors.
Under this point of view, of particular interest is the emerging evidence that at least  
some short
GRBs are followed by an extended, weak and soft emission. For one of the events included
in the \swift{} sample of short GRBs with known redshift, 
GRB\,050724, an estimate of the \eiso{} and \epi{} of this soft component is available
(see Table 1), thanks
to the joint fit of XRT and BAT data \cite{Barthelmy05}. Intriguingly, as can be seen
in  Figure 1, the extended soft emission of the short GRB\,050724 is fully consistent with
the \epeiso{} correlation. This evidence, if confirmed by future observations, 
may suggest that the emission mechanisms responsible for
most of the emission of long GRBs could also be at work in short GRBs but with a much
lower efficiency.

\begin{figure}
\centerline{\includegraphics[width=11cm]{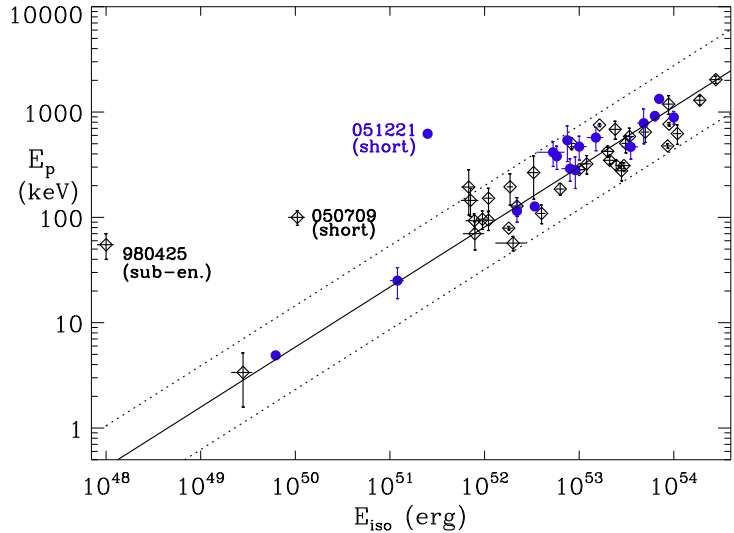}}
\caption{Location in the \epeiso{} plane of the most updated sample of GRBs with
known redshift and {\it accurate} estimate of \epi, including 50 long GRBs, 2 short GRBs
and the peculiar sub--energetic GRB\,980425. \swift{} GRBs are shown as filled circles.
The continuous line is the
power--law best fitting the \epeiso{} correlation and the dashed lines
delimitate the 2$\sigma$ confidence region (from \cite{Amati06}).}
\end{figure}

\section{Conclusions}

\swift, thanks to the combination of the high sensitivity of BAT with the few arcsec
source location accuracy of XRT and the very fast
slewing capability of the spacecraft, is making possible a substantial reduction of
selection effects in the sample of GRBs with known redshift, and thus
to test more stringently than in the past the \epeiso{} correlation.
As shown above, all \swift{} long GRBs with an estimate of \epi{}
(17 events) or 
an upper / lower limit to this quantity (5 events) are fully consistent
with the correlation. The results of the correlation analysis and power--law
fit of the these events,
which cover more than 4 orders of magnitude in \eiso{} and 3 orders of magnitude
in \epi, are fully consistent with what found for events detected by previous
satellites. This is a clear evidence, and a further confirmation, 
that the \epeiso{} correlation is likely not an artifact of selection effects.\\
Short \swift{} GRBs with known redshift (1 firm estimate of \epi{} and
5 lower limits) are inconsistent with the correlation, further confirming
that different emission mechanisms (possibly due to different conditions,
progenitors or circum--burst environment)
with respect to long GRBs are at work for this class of events.
Remarkably, the long, soft and weak tail following the short GRB\,050724
is characterized by values of \epi{} and \eiso{} fully consistent with the 
correlation holding for long GRBs, suggesting that
the emission mechanisms producing long GRBs could be at work also for
at least some short GRBs but with much less efficiency.
Finally, Figure 3 shows the location in the \epeiso{} plane of GRBs with known
redshift and more accurate estimates of \epi, a sample consisting of 51 long
GRBs plus 2 short GRBs. As can be seen, a part the two short
GRBs (050709 and 051221), only the peculiar and very close GRB\,980425
is a firm
outlier to the correlation.


\begin{thebibliography}{0}
%
\bibitem{Amati02} \BY{Amati L., Frontera F., Tavani M. et al.}
  \IN{A\&A}{390}{2002}{81}
%
\bibitem{Amati03} \BY{Amati L.}
  \IN{ChJAA}{3 Suppl.}{2003}{455}
%
\bibitem{Ghirlanda04a} \BY{Ghirlanda G., Ghisellini G. \atque Lazzati D.}
  \IN{ApJ}{616}{2004}{331}
%
\bibitem{Friedman05} \BY{Friedman A.S. \atque Bloom J.S.}
  \IN{ApJ}{627}{2005}{1}
%
\bibitem{Amati06} \BY{Amati L.}
  \IN{MNRAS}{372}{2006}{233}
%
\bibitem{Liang05} \BY{Liang E. \atque Zhang B.}
  \IN{ApJ}{633}{2005}{611}
%
\bibitem{Firmani06}\BY{Firmani C., Ghisellini G., Avila-Reese V., Ghirlanda G}
  \IN{MNRAS}{370}{2006}{185}
%
\bibitem{Nakar05} \BY{Nakar E. \atque Piran T.}
  \IN{MNRAS}{360}{2005}{L73}
%
\bibitem{Band05} \BY{Band D. \atque Preece R.}
  \IN{ApJ}{627}{2005}{319}
%
\bibitem{Ghirlanda05} \BY{Ghirlanda G., Ghisellini G. \atque  Firmani C. }
  \IN{MNRAS}{361}{2005}{L10}
%
\bibitem{Pizzichini06}\BY{Pizzichini et al.}
\IN{Adv.Sp.Res.}{38}{2006}{1338}
%
\bibitem{Band03}\BY{Band D.} 
\IN{ApJ}{588}{2003}{945}
%
\bibitem{Berger05}\BY{Berger E. et al.} 
\IN{ApJ}{634}{2005}{501}
%
\bibitem{Jakobsson06}\BY{Jakobsson P. et al.} 
\IN{A\&A}{447}{2006}{897}
%
\bibitem{Campana06}\BY{Campana S. et al.} 
\IN{Nature}{442}{2006}{1008}
%
\bibitem{Perri06}\BY{Perri M. et al.} 
\IN{A\&A}{submitted}{2006}
%
\bibitem{Mundell06}\BY{Mundell C.G. et al.} 
\IN{ApJ}{in press}{2006}{astro--ph/0610660}
%
\bibitem{Amati06b}\BY{Amati L. et al.} 
\IN{A\&A}{in press}{2006}{astro--ph/0607148}
%
\bibitem{Barthelmy05}\BY{Barthelmy S. et al.} 
\IN{Nature}{438}{2005}{994}
%


\end{thebibliography}
\end{document}